\documentclass[onecolumn,a4paper,aps,prd,floats,floatfix,amsmath,amssymb,amsfonts,nofootinbib,longbibliography,superscriptaddress,10pt]{revtex4}
\usepackage[utf8]{inputenc}
\usepackage[T1]{fontenc}
\usepackage{graphicx}
\usepackage{xcolor}
\usepackage{natbib}
\usepackage[colorlinks]{hyperref}
\newcommand\figref{Figure~\ref}

\begin{document}

%\title{A new formulation of a Newtonian-like gravity with variable $G$}

\title{Galactic core-tail structure in BEC dark matter with Kapitza potential}

\author{Itauany do Nascimento Barroso}
\email{itauany.barroso@edu.ufes.br}%
\affiliation{%
Departamento de Física, Universidade Federal do Espírito Santo (UFES), Campus Goiabeiras, 29.075-910, Vitória, Brasil}%

\author{Hermano Velten }
\email{hermano.velten@ufop.edu.br}%
\affiliation{% 
Departamento de Física, Universidade Federal de Ouro Preto (UFOP), Campus Morro do Cruzeiro, 35402-136, Ouro Preto, Brasil}%

\date{\today}

\begin{abstract}
Recently, the experimental realization of a Kapitza potential in a Bose-Einstein Condensate (BEC) has been reported for the first time in literature, motivating further theoretical investigations of such system. At the same time, in the astrophysical context, BEC dark matter models have been widely studied as a possible phenomenological explanation for the dark matter phenomena. We model the galactic structure with an inner cored profile obtained from the ground state equilibrium solution of the Schrödinger-Poisson together with a Kapitza-BEC like interaction for the tail region. We find reasonable agreement of the model with representative galaxy rotation curves available in the SPARC catalogue.

\end{abstract}


\maketitle

\section{Introduction}

The first evidence for dark matter (DM) emerged in 1933, when Fritz Zwicky studied the Coma Cluster and noticed that the visible mass was not sufficient to explain the observed dynamics \cite{zwicky1933}. 
Decades later, a different investigation route conducted by Vera Rubin and Kent Ford \cite{rubin1970rotation} confirmed this discrepancy, showing that the rotation curves of spiral galaxies remain approximately flat at large radial distances, in contrast with the Keplerian decay expected if only baryonic matter were present. 
These results paved the way that dark matter plays a fundamental role in the dynamics and formation of galaxies. Together with several other pieces of evidence from both astrophysical and cosmological contexts, dark matter became a fundamental ingredient in modern astronomy. See Refs. for reviews on the dark matter problem \cite{2024arXiv240601705C, BERTONE2005279, Bertone2018, feng2010dark}.

Widely used density profiles in the literature, such as Navarro-Frenk-White (NFW) \cite{Navarro1997} and Einasto \cite{Einasto1965} reproduce the observed behavior at intermediate and large galactocentric radii but present limitations in the central regions. While the NFW model predicts a steeply increasing central density (a cuspy profile), observations of dwarf and low surface brightness galaxies suggest the presence of cores with nearly constant density, motivating alternative profiles such as the Burkert model \cite{Burkert1995}.
This discrepancy is known as the cusp-core problem, and it remains a matter of debate in dark matter research \cite{de_Blok_2009,DelPopolo2017, David2015}.

In rotation curve analyses, different assumptions regarding the baryonic contribution are commonly adopted. One frequently used limiting case is the minimum disk assumption, in which the observed rotation curve is entirely attributed to the dark matter component. Although this approach is not physically realistic, since it neglects the contribution of the gaseous disk, which is observationally well established, it provides an upper bound on the allowed concentration of the dark halo. Moreover, it facilitates direct comparison with cosmological simulations, a large fraction of which model only the dark matter component \cite{de_Blok_high}. This assumption, therefore, provides a useful reference framework for assessing the behavior of dark matter dominated halo models.

Despite the strong evidence for the existence of a non-luminous component in the Universe, the nature of the dark matter particle remains unknown. 
Among the many theoretical candidates, an intriguing possibility is that dark matter could be composed of ultralight bosons, such as axions or axion-like particles \cite{marsh2016}. 
In this context, the large occupation number of these particles allows the formation of Bose-Einstein condensates (BECs) on astrophysical scales (see, for example, \cite{Bohmer2007, schive2014, hui2017}). This class of models is often referred to as fuzzy dark matter \cite{Hu:2000ke}, where the wave nature of ultralight bosons stabilizes central solitonic cores and mitigates the cusp-core discrepancy.

A BEC is a state of matter that arises when a system of bosons occupies the same quantum state at extremely low temperatures \cite{bose1924, einstein1925}, causing their collective behavior to be described by a single macroscopic wave function \cite{griffin1996bose} and theoretically described by the Gross-Pitaevskii equation \cite{gross1961, pitaevskii1961}. In laboratory settings, this phenomenon has been observed in dilute atomic gases cooled to temperatures near absolute zero \cite{anderson1995observation,  ketterle1999making}. In an astrophysical context, one can adopt the hypothesis that a similar process could occur naturally if dark matter were composed of ultralight bosonic particles with very small masses $m \sim 10^{-23}$ eV. In this regime, the wave nature of dark matter leads to the emergence of self-gravitating condensates whose dynamics are governed by the Gross-Pitaevskii-Poisson (GPP) system of equations.
Such models naturally predict core-tail halo structures, in which the central solitonic core is stabilized by quantum pressure, while the outer tail accounts for the approximately flat rotation curves observed at large radius \cite{schive2014, schive2014cosmic,alvarez2023}.

Furthermore, since the nature of DM and its possible self-interactions are still unknown, it is conceivable that it may be subject to alternative potentials, such as the Kapitza potential. Originally studied in the context of oscillating pendulums, the Kapitza potential produces a dynamic stabilization effect arising from rapid oscillations \cite{kapitza1951, He:2022kjp}. 
BEC systems trapped by the Kapitza potential have been recently realized in the laboratory for the first time \cite{jiang2021kapitza}. Hence, theoretical investigations of such systems are close to the state of the art in this field and, of course, welcomed.  
In the astrophysical scenario, introducing such an term can significantly modify the system’s dynamics, reshaping the density profiles in the tail region and offering a new perspective on halo stability and morphology. This approach also opens the possibility of reproducing the observed rotation curves with greater fidelity.

In this work, we qualitatively investigate the influence of the Kapitza potential parameters, particularly its amplitude, on the shape of rotation curves obtained from the modified GPP system. Our goal is to understand how variations in this parameter affect the halo stability, the transition radius between the core and tail, and the behavior of the orbital velocities. Apart from emphasizing a qualitative analysis aimed at revealing the physical role of the Kapitza potential in shaping galactic dynamics, we also employ a quantitative fitting with representative available data from the SPARC catalogue \cite{lelli2016}. For comparison purposes, we also confront the resulting total rotation curves with those obtained from standard density profiles widely used in the literature considering the full galactic rotation curve.

The paper is organized as follows: In the next section we present the model for the galactic structure and outline the governing equations. Section \ref{profiles} introduces the density profiles employed for comparison in the analysis, namely the NFW, Einasto, and Burkert models. Section \ref{results} presents the results, including the rotation curve fits, the corresponding density profiles, and an analysis of the influence of the interaction parameter $\epsilon$ and the Kapitza potential amplitude $V_{0\mathrm{Kap}}$. This section also includes a subsection with the best-fitting parameters and the chi-square values. In Section \ref{discussions} we provide a detailed discussion of the implications of our findings. Finally, Section \ref{conclusion} summarizes our conclusions.

\section{The model} \label{the model}

To describe our model, we start following the approach proposed in \cite{alvarez2023}. Modeling dark matter as a Bose-Einstein condensate requires the use of two equations.
The first is the Gross-Pitaevskii (GP) equation, a nonlinear variant of the time-dependent Schrödinger equation that governs the temporal evolution of the macroscopic wave function of the condensate.
This equation is employed because it effectively accounts for the interactions between identical particles in the condensate, and it is given by    
\begin{equation} \label{eqGPP}
    i \hbar \partial_t \psi = - \frac{\hbar^2}{2 m} \nabla^2 \psi + \frac{4 \pi \hbar^2}{m} a_s|\psi|^2 \psi + V_{tot}\psi,
\end{equation}
where $\hbar$ is the reduced Planck constant, $t$ is the temporal coordinate, $r$ is the radial coordinate, $m$ is the atomic mass, $a_s$ is the scattering length between atoms ($a_s > 0$ for repulsive interactions, $a_s < 0$ for attractive interactions), $V_{\text{tot}}$ is the total potential, and $|\psi|^2$ represents the particle density of the condensate. A subscript, such as $t$ or $x$, denotes a partial derivative with respect to that variable, that is,
$ \partial_t \psi \equiv \partial \psi / \partial t$.

The total potential $V_{\mathrm{tot}} = V + V_{\mathrm{Kap}}$ is composed of the gravitational potential $V$, which obeys the Poisson equation
\begin{equation}
\nabla^2 V = 4\pi G |\psi|^2,
\end{equation}
and by the potential $V_{\mathrm{Kap}}$ of the Kapitza type
\begin{equation}
V_{\mathrm{Kap}}(r,t)= V_{0\mathrm{Kap}} \exp\left(-\frac{2 r^2}{ a^2}\right) \,\cos{(\omega t + \phi_0)}.
\end{equation}
Here, $G$ in the Poisson equation is the gravitational constant. The parameters $a$ and $V_{0\mathrm{Kap}}$ are the waist and amplitude of the Gaussian profile, respectively, $\omega$ is the frequency and $\phi_0$ is the phase term of the oscillation.

\subsection{Madelung's transformation}

In order to reformulate the Gross-Pitaevskii equation in a hydrodynamic framework, we apply the Madelung transformation by writing the condensate wave function as $\psi = \sqrt{\rho}e^{iS}$, where $\rho$ is the mass density and $S$ is the phase of the wave function.  
This transformation enables us to describe the system using two real variables, $\rho$ and $S$, and interpret its dynamics within the quantum hydrodynamic formalism. Readers not familiar with the Madelung transformation and its physical interpretation may consult Refs. \cite{Madelung1927, Bohm1952}.
Substituting this expression into the Gross-Pitaevskii and Poisson equations and separating the real and imaginary parts, we obtain the following set of equations.

The real part of the Gross-Pitaevskii equation reads
\begin{equation} \label{p.real}
  \hbar \partial_t S + Q + \frac{\hbar^2}{2 m}|\vec{v}|^2 + \frac{4 \pi \hbar^2 a_s}{m} \rho + V + V_{0\mathrm{Kap}} \exp\left(-\frac{2 r^2}{a^2}\right)   \,\cos{(\omega t + \phi_0)} = 0,
\end{equation} 
the imaginary part yields the continuity equation
\begin{equation} \label{p.imaginária}
    \hbar \partial_t \rho + \frac{\hbar^2}{m} \left[\nabla \cdot (\rho \vec{v}) \right] = 0,
\end{equation} 
and the Poisson equation becomes,
\begin{equation} \label{poisson2}
     \nabla^2 V = 4 \pi G\rho.
\end{equation} 

Here, $\vec{v} := \nabla S$ represents the velocity field of the quantum fluid. The definition
\begin{equation}
   Q = -\frac{\hbar^2}{2 m}\frac{\nabla^2 \sqrt{\rho}}{\sqrt{\rho}} 
\end{equation}
represents the quantum potential, and $V$ is the gravitational potential generated by the mass density $\rho$.  
This hydrodynamic formulation is widely adopted in the literature because it highlights the analogy between a self-gravitating condensate and a compressible fluid with an additional quantum pressure term. For convenience, the notation can be simplified by adopting natural units, $\hbar = m = 4\pi G = 1$ and we define $\epsilon = 4 \pi a_s$.
Assuming spherical symmetry and a stationary density profile, we take the ansatz for the phase
\begin{equation}
    S(t, r) = -V_0 t + \overline{S}(r),
\end{equation}
where $r$ is the radial coordinate.  Since we seek stationary solutions, the Kapitza potential is treated as time independent by setting $t=0$ and 
$\phi_0 = 0$, so that only its radial dependence enters the equations. This assumption allows us to isolate the static influence of the Kapitza term on the tail structure. Under the assumptions of stationarity and spherical symmetry, the Gross-Pitaevskii-Poisson system reduces to three coupled ordinary differential equations
\begin{equation} \label{p.real3}
     Q + \frac{1}{2} v^2 + \epsilon \rho  + V + V_{0\mathrm{Kap}} \exp\left(-\frac{2 r^2}{a^2}\right) = V_0,
\end{equation}  
\begin{equation} \label{p.imaginária3}
    \frac{1}{r^2} \frac{d}{dr} (r^2 \rho v) = 0,
\end{equation}
\begin{equation} \label{poisson4}
     \frac{1}{r^2} \frac{d}{dr} \left(r^2 \frac{dV}{dr}\right) = \rho,
\end{equation} 
where the radial velocity is $v := \frac{d \overline{S}}{dr}$. 

\subsection{Core and tail structure}

In order to construct a consistent description of the dark matter halo, we adopt a core-tail decomposition, in which the system is divided into two radial regions separated by a transition radius $r_t$. The inner region, referred to as the core, corresponds to $r<r_t$ and is characterized by a solitonic configuration, while the outer region, referred to as the tail, corresponds to $r\geq r_t$ and is described by the stationary solution of the Gross-Pitaevskii-Poisson system. 

Solving the system \eqref{p.real3}-\eqref{poisson4}, we find
\begin{equation} \label{system-rho}
   \rho'_{\text{tail}} = u_{\text{tail}},
\end{equation}
% \begin{equation} \label{system-u}
% \begin{split}
%    u'_{\text{tail}} & = 4\left[\frac{1}{2}v_{\text{tail}}^2 + V_{\text{tail}} - V_0 
%    + V_{0\,\text{kap}} \exp\left(-\frac{2r^2}{a^2}\right)\cos(\omega t + \phi_0)\right]\rho_{\text{tail}} \\
%    & \quad - \frac{2u_{\text{tail}}}{r} + \frac{u_{\text{tail}}^2}{2\rho_{\text{tail}}} 
%    + 4\epsilon \rho_{\text{tail}}^2,
% \end{split}
% \end{equation}
\begin{equation} \label{system-u}
   u'_{\text{tail}} = 4\left[\frac{1}{2}v_{\text{tail}}^2 + V_{\text{tail}} - V_0 
   + V_{0\mathrm{Kap}} \exp\left(-\frac{2r^2}{a^2}\right)\right]\rho_{\text{tail}} - \frac{2u_{\text{tail}}}{r} + \frac{u_{\text{tail}}^2}{2\rho_{\text{tail}}} 
   + 4\epsilon \rho_{\text{tail}}^2,
\end{equation}
\begin{equation} \label{system-V}
   V'_{\text{tail}}(r) = \frac{M_{\text{tail}}(r)}{r^2},
\end{equation}
\begin{equation} \label{system-M}
   M'_{\text{tail}}(r) = r^2 \rho_{\text{tail}}(r),
\end{equation}
\begin{equation} \label{system-S}
   \overline{S}'_{\text{tail}}(r) = v_{\text{tail}}(r),
\end{equation}
\noindent
where $V_0$ is constant, $' := \frac{d}{dr}$ and $v_{\text{tail}} = \frac{A}{r^2 \rho_{\text{tail}}}$ is the velocity in the tail region, with $A$ constant. In the absence of self-interactions and of the Kapitza potential, i.e., for $\epsilon = 0$ and $V_{0\mathrm{Kap}} = 0$, the equations obtained here reduce to those in Ref.~\cite{alvarez2023}, which provides the reference framework upon which the present model is built.

In the innermost region of the halo, the density profile is dominated by a solitonic structure corresponding to the ground state equilibrium solution of the Gross-Pitaevskii-Poisson system. In the non-interacting limit, this system reduces to the Schrödinger-Poisson equations solved in Refs.~\cite{schive2014cosmic, schive2014}. Using fully cosmological, high-resolution three-dimensional simulations, the authors of the latter references showed that every virialized halo hosts a central, gravitationally self-bound solitonic core. By fitting the numerically obtained soliton solutions, an accurate analytical approximation for the core density profile was derived, valid within a few core radii. The resulting core density profile is given by
\begin{equation} \label{core-density}
   \rho_{\text{core}}(r) = \rho_{0 {\text{core}}} 
   \left[1 + 0.091 \left(\frac{r}{r_c}\right)^2 \right]^{-8},
\end{equation}
where $\rho_{0\text{core}}$ is the central density and $r_c$ is the characteristic core radius.

Since the radial derivative of the density is defined as $u_{\text{core}} = \rho'_{\text{core}}$, we obtain
\begin{equation} \label{core-u}
   u_{\text{core}}(r) = 
   -\frac{1.456\,r \, \rho_{0 {\text{core}}}}{\left(1 + \frac{0.091 r^2}{r_c^2}\right)^9 r_c^2}.
\end{equation}

The total mass enclosed within radius $r$ is given by
\begin{equation} \label{core-mass}
   M_{\text{core}}(r) = 
   \int_0^{r} \rho_{\text{core}}(r')\, r'^2\, dr',
\end{equation}
and the corresponding gravitational potential follows from Poisson’s equation
\begin{equation} \label{core-potential}
   V_{\text{core}}(r) = 
   -\,\frac{GM_{\text{core}}(r)}{r}.
\end{equation}

The central structure, being spherically symmetrical and dynamically stable, represents the region of the halo sustained by quantum pressure. It transitions smoothly into the outer tail, where the density decreases gradually and the classical regime becomes dominant.

To ensure continuity between the core and
the tail, we impose the following boundary conditions:
\begin{center}
    $\rho_{\text{tail}}(r_t) = \rho_{\text{core}} (r_t)$, \\
    $ u_{\text{tail}}(r_t) = u_{\text{core}} (r_t)$, \\
    $ V_{\text{tail}}(r_t) = V_{\text{core}} (r_t)$, \\
    $ M_{\text{tail}}(r_t) = M_{\text{core}} (r_t)$. 
\end{center}

\section{Dark matter density profiles} \label{profiles}

In this section, we present the dark matter density profiles widely used in the literature that will be used as reference models for comparison with the results obtained from the model developed in this paper.

\subsection{Navarro-Frenk-White (NFW) Profile}

The NFW density profile \cite{Navarro1997} is defined by
\begin{equation}
    \rho_{\mathrm{NFW}}(r) = 
    \frac{\rho_s}{\left( \frac{r}{r_s} \right)\left(1 + \frac{r}{r_s}\right)^2},
\end{equation}
where $\rho_s$ is a characteristic density and $r_s$ is the scale radius.

The enclosed mass has an analytical solution:
\begin{equation}
    M_{\mathrm{NFW}}(r) =
    4\pi\rho_s r_s^3
    \left[
        \ln(1+x) - \frac{x}{1+x}
    \right],
    \qquad x=\frac{r}{r_s}.
\end{equation}

% The circular velocity is:
% \begin{equation}
%     v_{\mathrm{NFW}}(r)
%     =
%     \sqrt{\frac{GM_{\mathrm{NFW}}(r)}{r}}.
% \end{equation}

\subsection{Einasto Profile}

The Einasto density profile was defined in the paper \cite{Einasto1965}. In this work, we adopted the approach \cite{Merritt_2006}, where
\begin{equation}
    \rho_{\mathrm{Ein}}(r)
    =
    \rho_{-2}\,
    \exp\left[
        -\frac{2}{\alpha}
        \left(
            \left(\frac{r}{r_{-2}}\right)^{\alpha} - 1
        \right)
    \right].
\end{equation}
Here, $\rho_{-2}$ and $r_{-2}$ are the density and radius at which the logarithmic slope of the density profile is equal to $-2$, i.e. $\left. d\ln\rho / d\ln r \right|_{r=r_{-2}} = -2$. For details on other conventions and their corresponding conversions, see \cite{Retana_Montenegro_2012}.
%(XXX For other conventions and the conversion between them, we refer to RetanaMontenegro et al. (2012a).)

The enclosed mass is given by $ M_{\mathrm{Ein}}(r)=
    4\pi\int_{0}^{r}
    \rho_{\mathrm{Ein}}(r')\,r'^2\,dr'.$
    
\subsection{Burkert Profile}

The Burkert density profile \cite{Burkert1995} is defined as
\begin{equation}
    \rho_{\mathrm{Bur}}(r)
    =
    \frac{\rho_0}{
        \left(1 + \frac{r}{r_0}\right)
        \left[1 + \left(\frac{r}{r_0}\right)^2\right]
    },
\end{equation}
where $\rho_0$ is the central density and $r_0$ is the scale radius.

The enclosed mass has an analytical form given by
\begin{equation}
    M_{\mathrm{Bur}}(r)
    =
    \pi\rho_0 r_0^3
    \left[
        \ln(1+x^2)
        +2\ln(1+x)
        -2\arctan(x)
    \right],
    \qquad x=\frac{r}{r_0}.
\end{equation}

% The circular velocity is:
% \begin{equation}
%     v_{\mathrm{Bur}}(r)
%     =
%     \sqrt{\frac{GM_{\mathrm{Bur}}(r)}{r}}.
% \end{equation}

\section{Results} \label{results}

To adjust our model with observational data, transforming the system into physical units, we use scaling factors for time and mass, respectively:
\begin{equation}
    t_S = \frac{m r_S^2}{\hbar}, \,\,\, M_S = \frac{\hbar^2}{4\pi G m^2 r_S}.
\end{equation}
The scale factors are defined from $r_S$, which represents a specific length chosen according to  Ref. \cite{alvarez2023}.  From these quantities, other scaling factors can be defined, such as the density scale $\rho_S = M_S/r_S^3$ and the velocity scale $v_S = r_S/t_S$.

\begin{figure}[!ht]
\centering
\includegraphics[width=.48\textwidth]{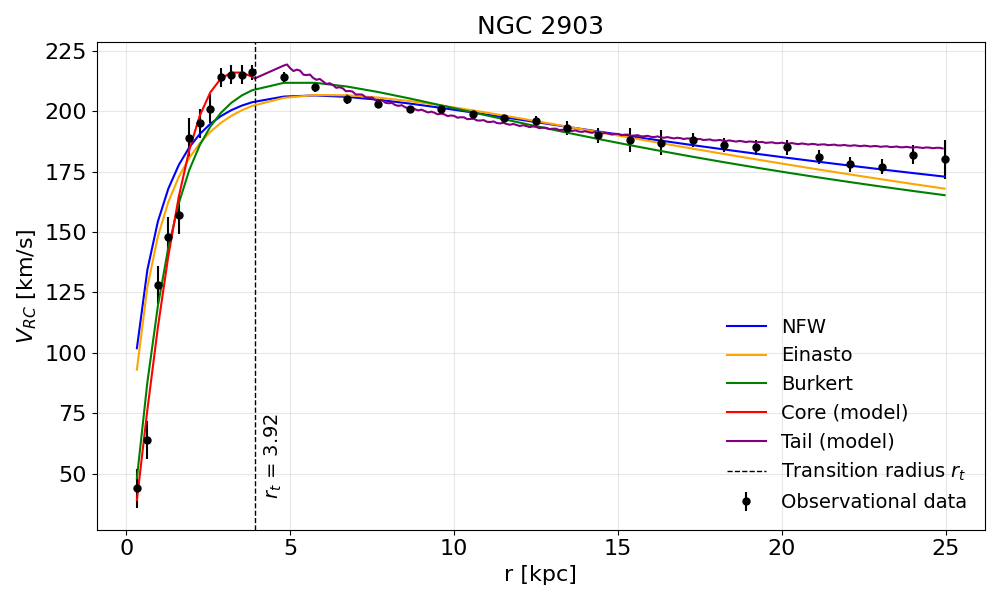} % end subfigure
\quad \includegraphics[width=.49\textwidth]{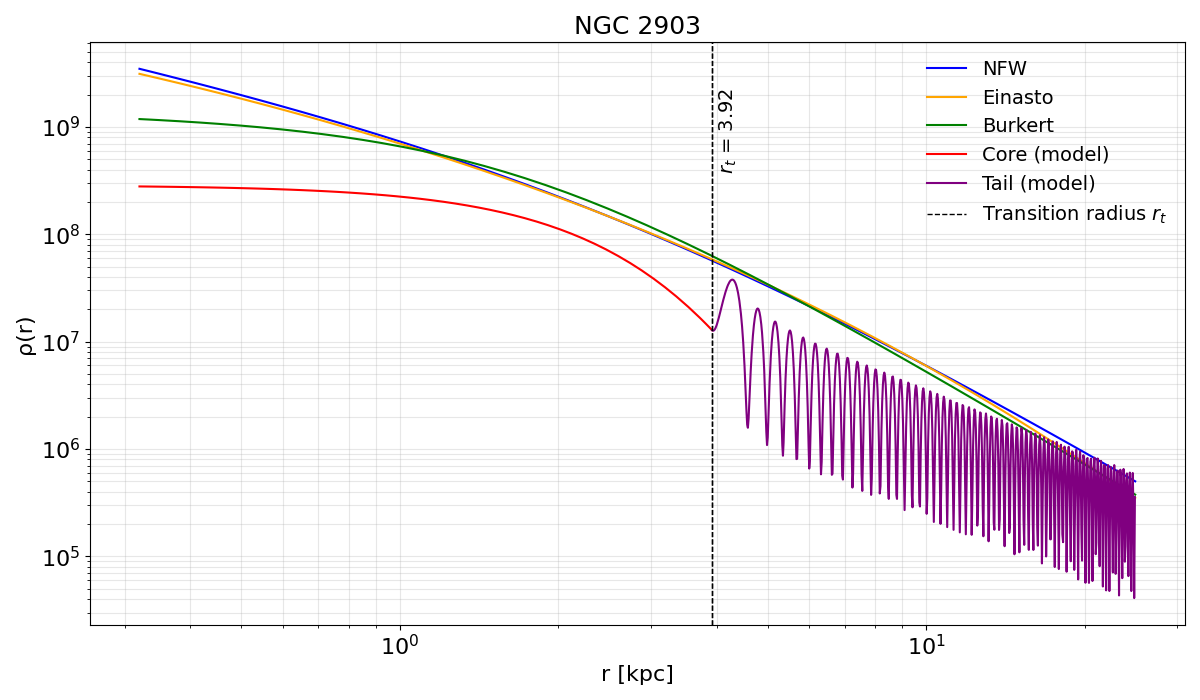} % end subfigure
\quad % dá um espaço entre as duas figuras.
\includegraphics[width=.48\textwidth]{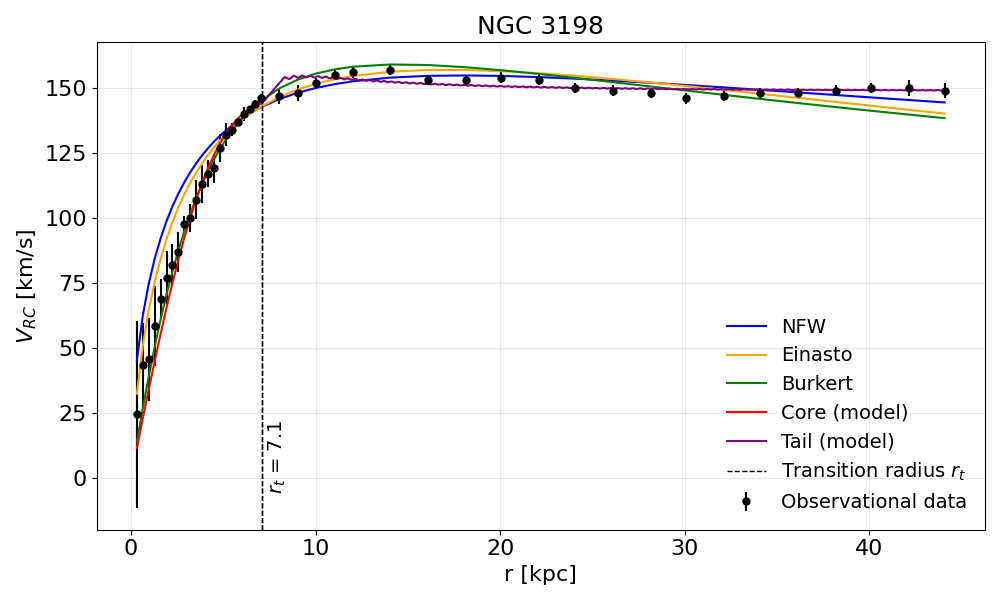} % end subfigure
\quad \includegraphics[width=.49\textwidth]{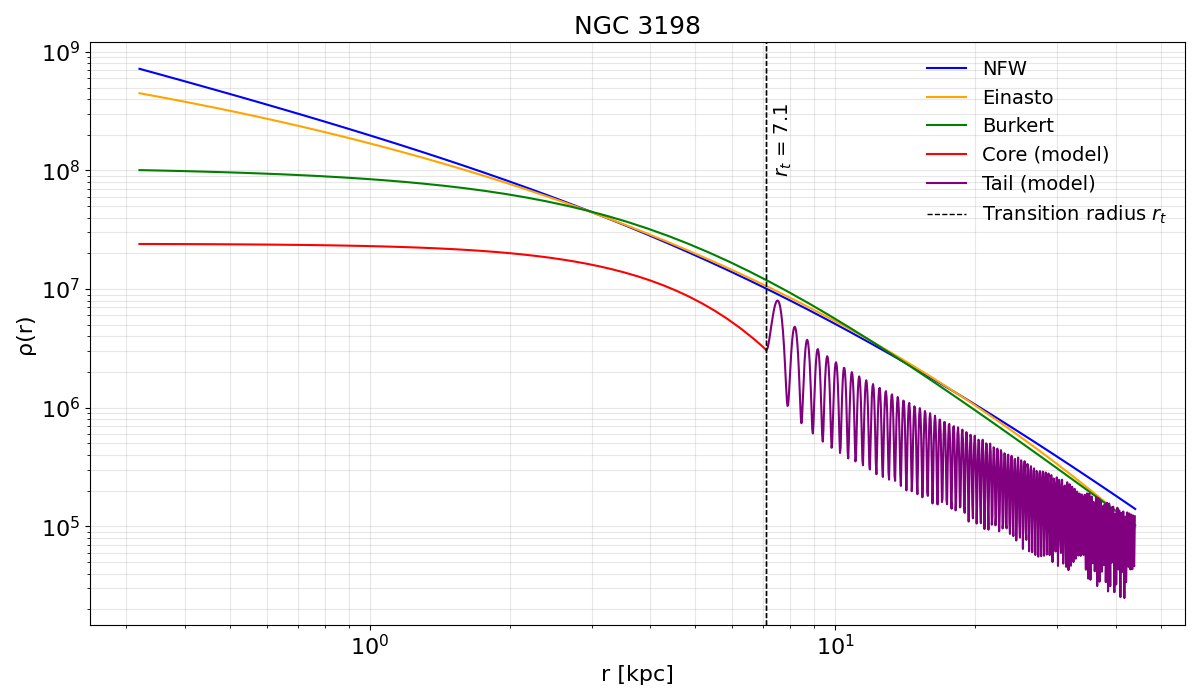} % end subfigure
\quad % dá um espaço entre as duas figuras.
\includegraphics[width=.48\textwidth]{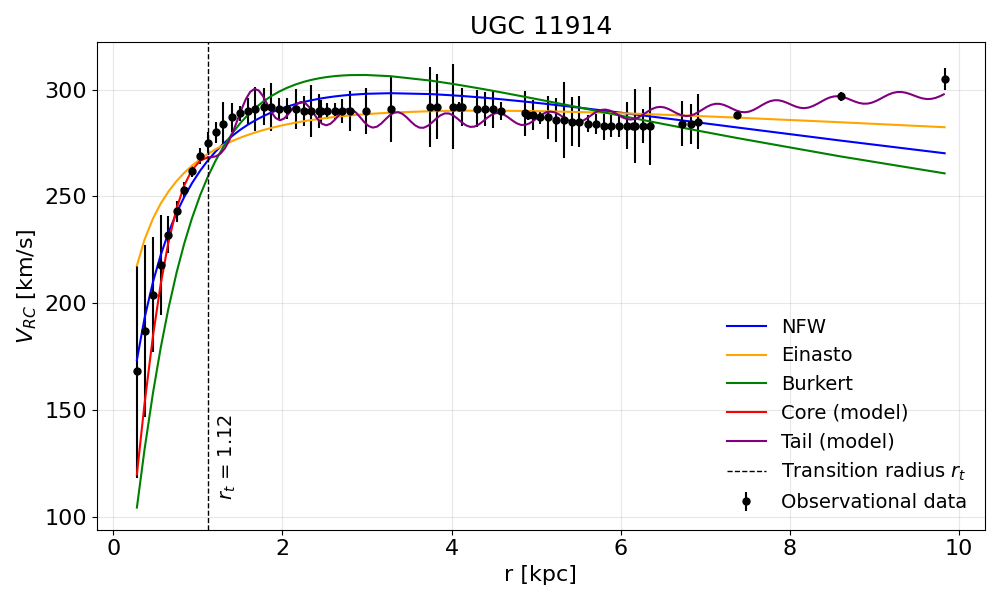} % end subfigure
\quad % dá um espaço entre as duas figuras.
\includegraphics[width=.49\textwidth]{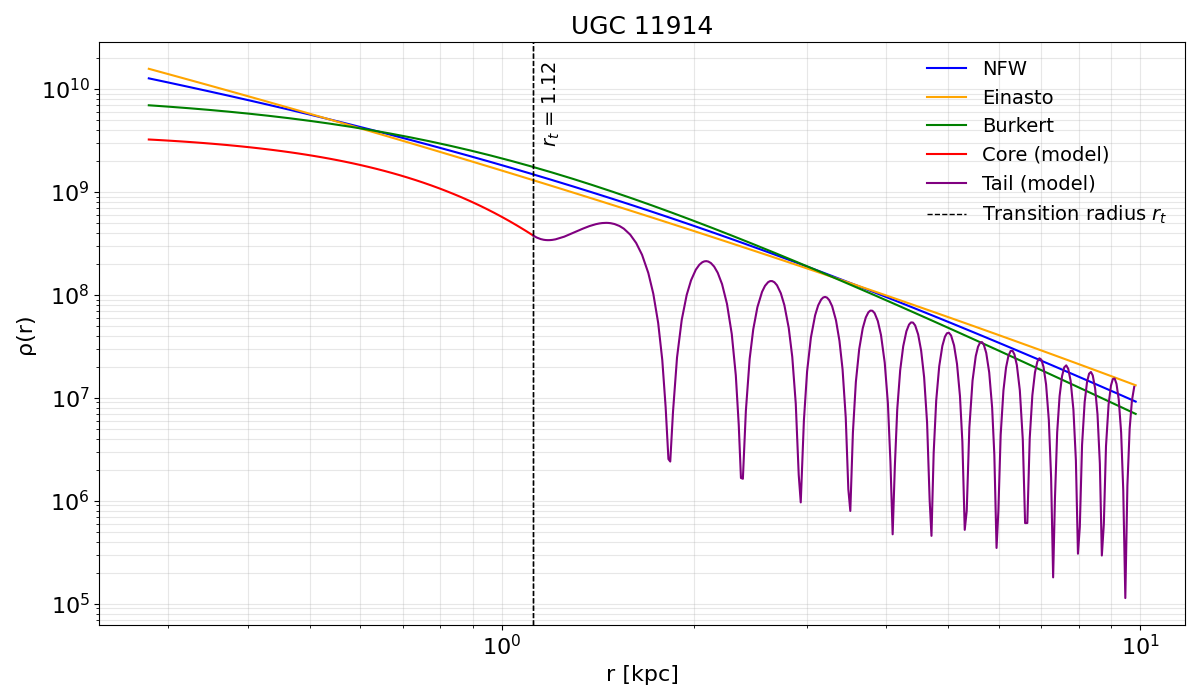}
\quad 
\includegraphics[width=.48\textwidth]{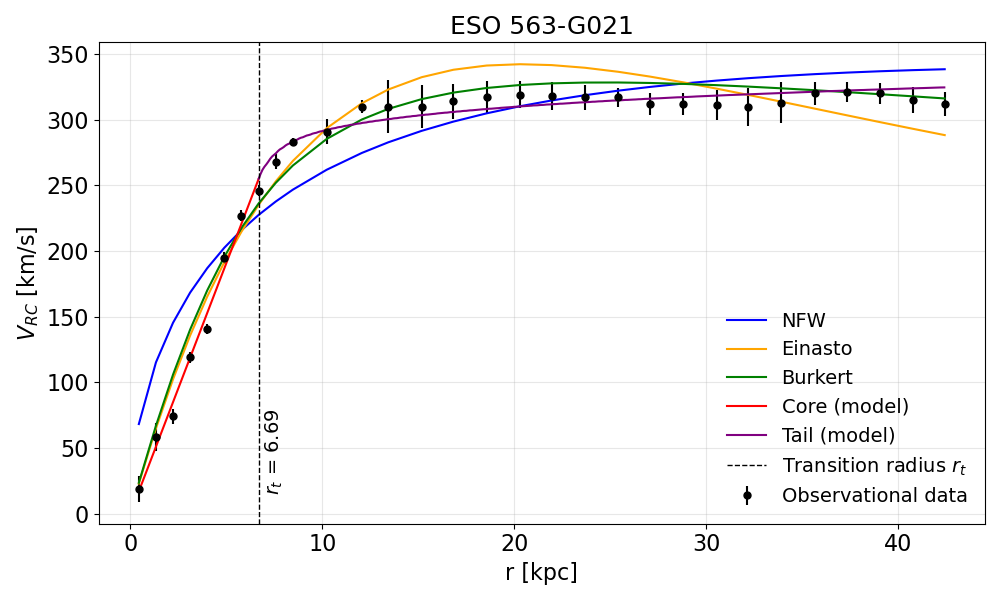} % end subfigure
\quad
\includegraphics[width=.49\textwidth]{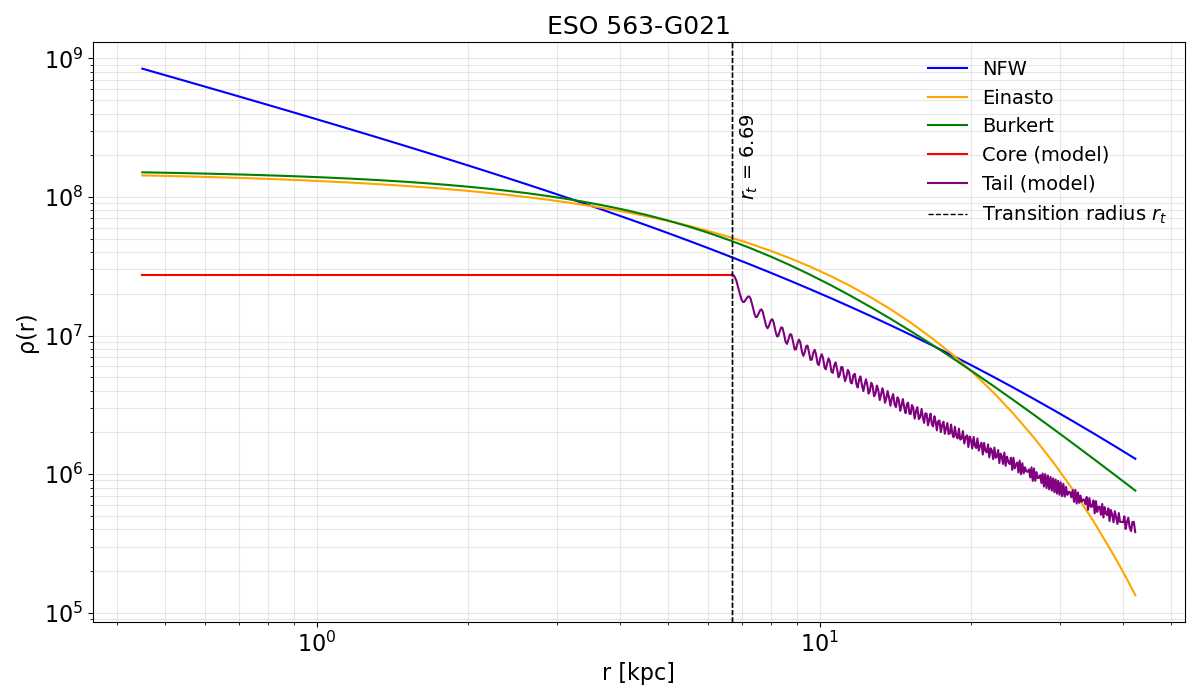} % end subfigure
\caption{Rotation curves (left panels) and density profiles in log-log scale (right panels) for the galaxies analyzed. The NFW, Einasto, Burkert profiles and the solitonic core-tail model are shown. The dashed line indicates the transition radius $r_t$. The adjusted parameters are listed in Table \ref{table1}.}
\label{galaxies1}
\end{figure}

\begin{figure}[!htb]
\centering
% dá um espaço entre as duas figuras.
\includegraphics[width=.48\textwidth]{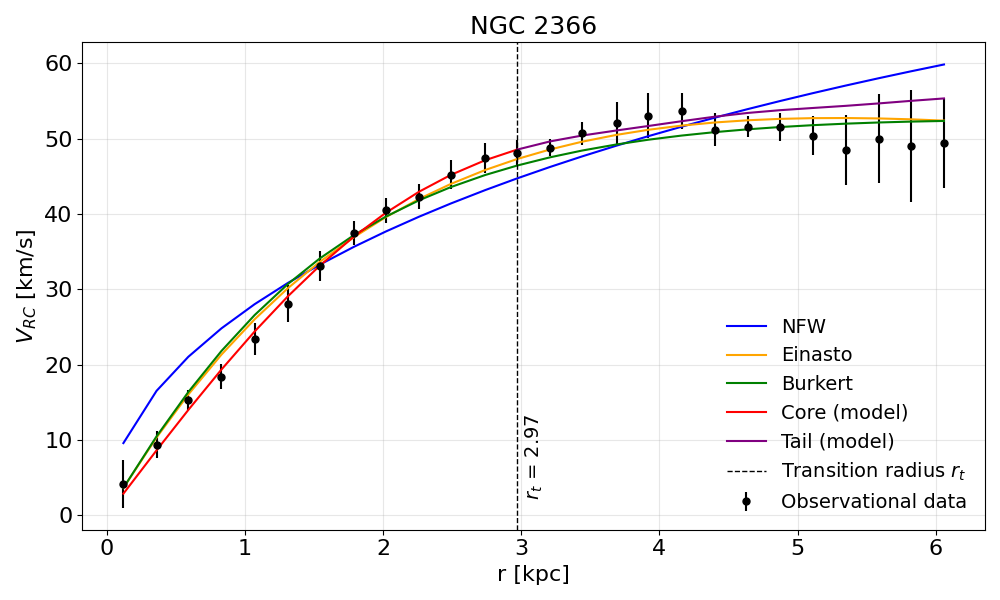} % end subfigure
\quad \includegraphics[width=.49\textwidth]{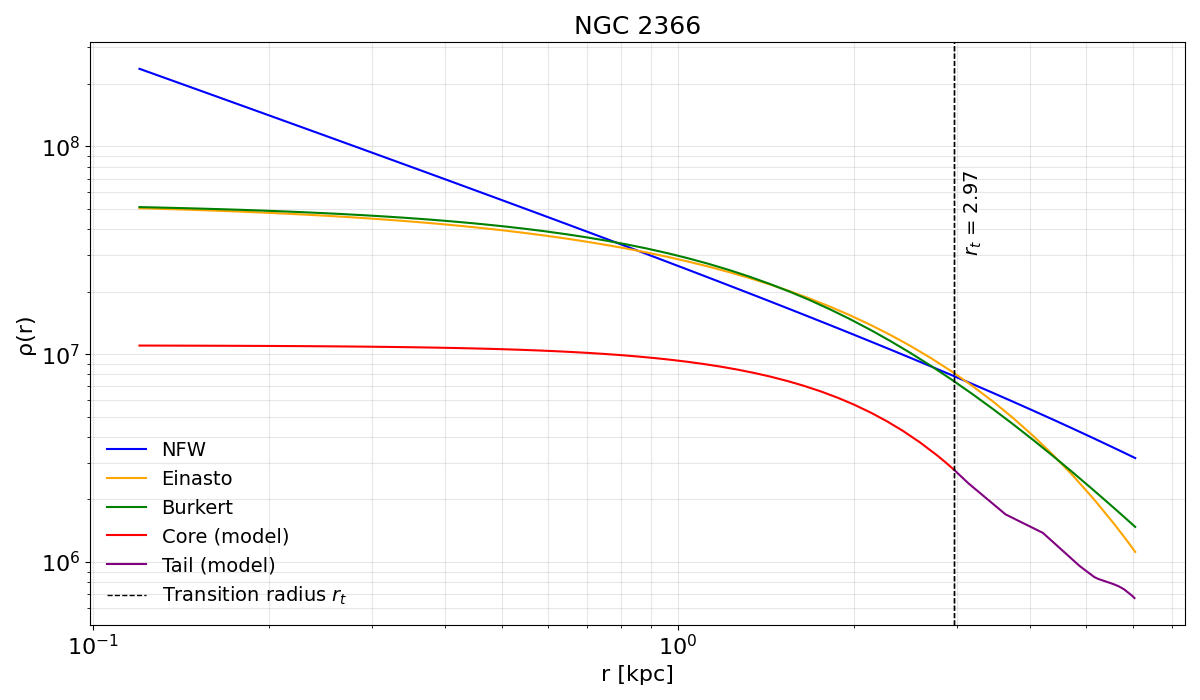} % end subfigure
\quad % dá um espaço entre as duas figuras.
\includegraphics[width=.48\textwidth]{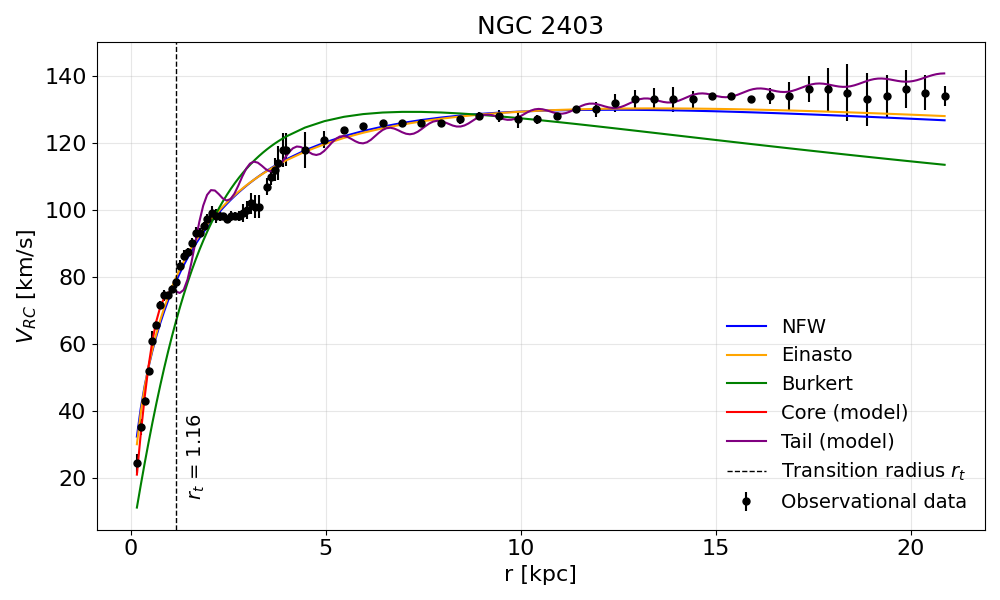}
\includegraphics[width=.49\textwidth]{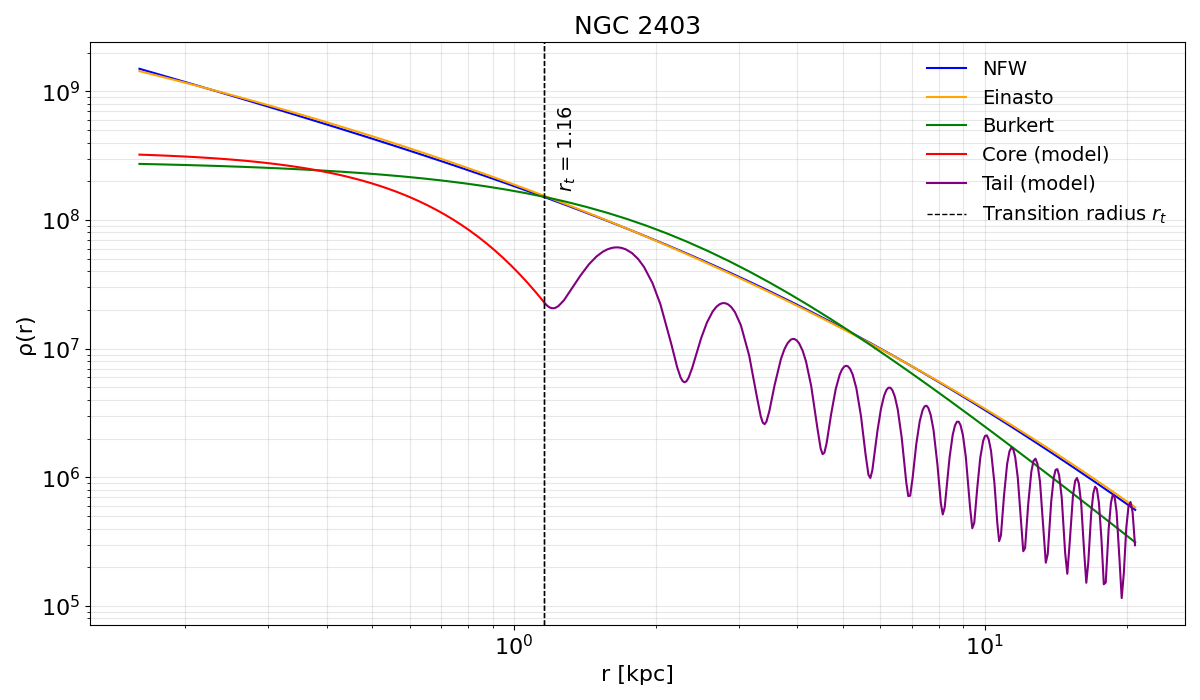}
\quad
\includegraphics[width=.48\textwidth]{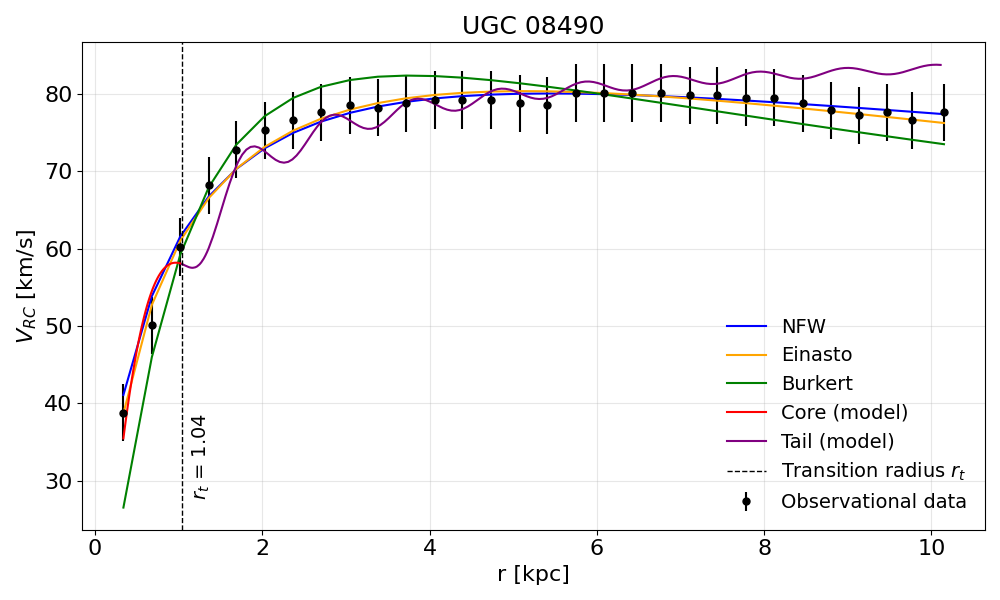} % end subfigure
\quad
\includegraphics[width=.49\textwidth]{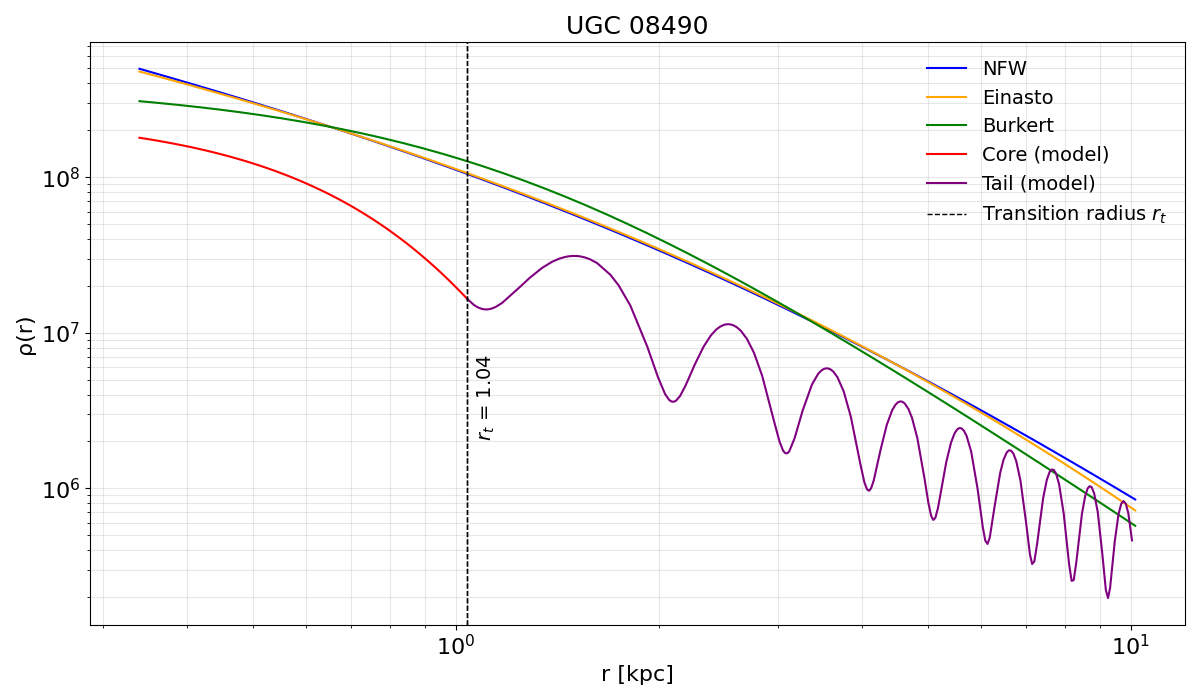} % end subfigure
\caption{Same as \figref{galaxies1} for different galaxies.}
\label{galaxies2}
\end{figure}

To work with typical galaxy scales, let us consider a reference wavelength $r_S = 1.714$ kpc and an ultra-light boson mass
$m = 10^{-23} \mathrm{eV/c}^{2}$. From these values, the scaling factors for time and mass are given by
$t_S = 14.97$ Myr and $M_S = \frac{5}{4\pi} \times {10^{9}} M\odot$. Furthermore, the scale factors associated with density and velocity are
$\rho_S = 7.901 \times 10^{7} \mathrm{M}\odot/\mathrm{kpc}^{3}$ and
$v_S = 112.0$ km/s, respectively. Thus, to obtain the rotation curves and density profiles of our model in physical units, we multiply the scaling factors $v_S$ and $\rho_S$ in the final expressions for velocity and density, respectively.

\subsection{Rotation Curves and Density Profiles} 

To plot the rotation curves (RC) for all density models considered in this work, we adopt the general expression
\begin{equation}
    v_{\mathrm{RC}}(r)
    =
    \sqrt{\frac{GM(r)}{r}},
\end{equation}
where $M(r)$ is the enclosed mass associated with the corresponding density profile, whether defined analytically (NFW, Einasto, Burkert, solitonic core) or obtained numerically from the system in the tail region. We have therefore a kind of hybrid model, both analytical for the solitonic core and numerical for the tail. 

\figref{galaxies1} and \figref{galaxies2} present the rotation curves (left panels) resulting from the corresponding density profiles (right panels) for the galaxies NGC 2903, NGC 3198, UGC 11914, ESO 563-G021, NGC 2366, NGC 2403 and UGC 08490. In the left panels one can also visualize the observational data with the fits obtained by the different models tested. The adopted parameters values are shown later in \ref{table1}. Within the adopted minimum disk approach, each classical profile (NFW, Einasto, and Burkert) is contrasted with the model proposed in this work, composed of a solitonic core and a dynamic tail region. The transition radius $r_t$, which delimits the interface between the core and tail, is indicated by a vertical dashed line in all panels. The chosen values of $r_t$ were selected to coincide with the onset of flatness behavior. We have also checked that our final conclusions depend weakly on the choice for the specific $r_t$ value around the chosen ones. The rotation curves show the predicted behavior for the circular velocity $V_{RC}(r)$, while the log-log plots display the radial density drop for each profile. 

\subsection{Fitting analysis}

To adjust the NFW, Einasto, Burkert, and soliton core density profiles, we used the nonlinear least squares method implemented in Python through the \textit{curve\_fit} algorithm from the \textit{scipy.optimize} module \cite{Virtanen2020}. For the tail region, which is described by a dynamical model derived from a system of coupled differential equations, a different strategy was adopted. Instead of a direct functional fit, the tail solution was obtained through numerical integration of the system for each combination of the free parameters  $A$, $V_0$ and $V_{0\mathrm{Kap}}$. For a given parameter set, the resulting rotation curve was evaluated at the same radii as the observational data, and the quality of the fit was quantified using the chi-square statistic. The optimal parameters were identified through a systematic exploration of the parameter space, seeking solutions that minimize chi-square statistics which is described below.

The chi-square method ($\chi^{2}$) allows us to quantify the deviation between the values predicted by the model and the observed velocities, weighting each discrepancy by the experimental uncertainties associated with each data point (books such as \cite{2003drea.book.....B} offer in-depth explanations of the method). Explicitly, the chi-square is defined by
\begin{equation}
\chi^2 = \sum_{i=1}^{N} \frac{\left[ v_{\text{obs}}(r_i) - v_{\text{mod}}(r_i) \right]^2}{\sigma_i^2},
\end{equation}
where $v_{\text{obs}}(r_i)$ is the observational velocity at radius $r_i$, $v_{\text{mod}}(r_i)$ is the velocity calculated by the model at the same point, and $\sigma_i$ represents the error associated with the experimental measurement. Thus, smaller values of $\chi^2$ correspond to models that provide a closer fit to the observed rotation curves.

% {\bf XXX( Talvez não seja necessário falar do chi reduzido)} Although the absolute value of $\chi^2$ is informative, we use the reduced chi-square test to determine the adjusted tail parameters, as this metric normalizes the error by the number of degrees of freedom and allows us to reliably identify which parameter combinations produce statistically consistent fits. Thus, the reduced chi-square is defined as
%\begin{equation}
%\chi_{\nu}^{2} = \frac{\chi^2}{N - p},
%\end{equation}
%where $N$ is the number of observational points and $p$ is the number of adjusted parameters. This procedure normalizes the chi-square by the number of degrees of freedom, $\nu = N - p$, allowing for consistent statistical interpretation. Values of $\chi_{\nu}^{2} \approx 1$ indicate that the model adequately reproduces the observed velocities within the experimental uncertainties; much higher values suggest physical inadequacy or unconsidered systematic errors, while much lower values may indicate overfitting or underestimation of uncertainties. (XXX CORTAR ATÉ AQUI)

Table \ref{table1} presents the adjustments obtained for the different density profiles applied to the analyzed galaxies. We have fixed the value parameters $\epsilon = 1$ and $a = r_t$ in all fits.
 For each object, the adjusted parameters of the model proposed in this work are listed, as well as the characteristic parameters of the NFW, Einasto, and Burkert profiles. The table also shows the respective chi-square values $\chi^2$, allowing comparison of the performance of each model in reproducing the observed rotation curves.

\begin{table}
\centering
\caption{Best-fitting parameters of the profiles. Fixed parameters: $\epsilon = 1$, $a = r_t$.}
\begin{tabular}{cccc}
\hline
\textbf{Galaxy}   & \textbf{Model}   & \textbf{Parameters adjusted} & \textbf{$\chi^2$} 
\\ \hline
NGC 2903 & Our model    &  \begin{tabular}[c]{@{}c@{}}$ \rho_{\text{core}} = 3.63148$, $r_c = 1.71602$,\\  $A= 11.5172$, $V_0 = 50.9184$, $V_{0\mathrm{Kap}} = 466.9388$     \end{tabular}               &     56.5571        \\
NGC 2903 & NFW     &   $\rho_s  = 5.397 \cdot 10^8$, $r_s = 2.60$            &      248.318      \\
NGC 2903 & Einasto &  $\rho_{-2} = 1.069 \cdot 10^8 $, $r_{-2} = 2.93$, $\alpha = 0.258$                   &    268.458         \\
NGC 2903 & Burkert &  $\rho_0 =  1.472 \cdot 10^9 $, $r_0 = 1.62$         &  175.458
\\ \hline
NGC 3198 & Our model    &  \begin{tabular}[c]{@{}c@{}} $\rho_{\text{core}} = 0.30623$, $r_c = 3.95455$,\\ $A = 9.7960$, $V_0 = 31.7347$, $V_{0\mathrm{Kap}} = 243.6735$\end{tabular}          &   44.1834         \\
NGC 3198 & NFW     &     $\rho_s  = 3.006 \cdot 10^7$, $r_s = 8.26$                    &  145.242           \\
NGC 3198 & Einasto &   $\rho_{-2} =7.63 \cdot 10^6 $, $r_{-2} = 8.42$, $\alpha = 0.317$                   &  129.295            \\
NGC 3198 & Burkert &  $\rho_0 = 1.084 \cdot 10^8 $, $r_0 = 4.48$                     & 104.409
\\ \hline
UGC 11914 & Our model    & \begin{tabular}[c]{@{}c@{}}$ \rho_{\text{core}} = 48.47837$, $r_c = 0.58303$, \\   $A = 6.0606$, $V_0 = 17.1429$, $V_{0\mathrm{Kap}} = 263.6364$,  \end{tabular}                  &   56.6016          \\
UGC 11914 & NFW     &    $\rho_s  =  3.254 \cdot 10^9$, $r_s = 1.53$              &    254.756          \\
UGC 11914 & Einasto &  $\rho_{-2} = 4.927 \cdot 10^8 $, $r_{-2} = 1.84$, $\alpha = 0.09$                &  147.617           \\
UGC 11914 & Burkert &  $\rho_0 = 1 \cdot 10^{10} $, $r_0 = 0.90$          &  686.048 
\\ \hline
ESO 563-G021 & Our model    & \begin{tabular}[c]{@{}c@{}}$ \rho_{\text{core}} = 0.34854$, $r_c = 70439.62462$, \\ $A = 81.6327$, $V_0 = 46.5306$, $V_{0\mathrm{Kap}} = 265.9184$        \end{tabular}                &      39.6443       \\ 
ESO 563-G021 & NFW     &    $\rho_s  =  1.559 \cdot 10^7$, $r_s = 25.22$              &    811.634         \\
ESO 563-G021 & Einasto &  $\rho_{-2} = 2.087 \cdot 10^7 $, $r_{-2} = 12.03$, $\alpha = 1$                &  198.959          \\
ESO 563-G021 & Burkert &  $\rho_0 = 1.601 \cdot 10^{8} $, $r_0 = 7.61$          &  192.979
\\ \hline
NGC 2366 & Our model    &  \begin{tabular}[c]{@{}c@{}}$ \rho_{\text{core}} = 0.14003$, $r_c = 2.06459$, \\  $A = 0.6122$, $V_0 = 2.0408$, $V_{0\mathrm{Kap}} = 1.0204$ \end{tabular}                 &   13.2718          \\
NGC 2366 & NFW     &    $\rho_s  =  1.045 \cdot 10^6$, $r_s = 27.37$              &   110.4         \\
NGC 2366 & Einasto &  $\rho_{-2} = 7.366 \cdot 10^6 $, $r_{-2} = 3.12$, $\alpha = 1$                &  13.1045           \\
NGC 2366 & Burkert &  $\rho_0 = 5.41486 \cdot 10^7 $, $r_0 = 2.08978$          &  21.6403
\\ \hline
NGC 2403 & Our model    &  \begin{tabular}[c]{@{}c@{}} $\rho_{\text{core}} = 4.34883$, $r_c = 0.550723$\\ $A = 2.6531$, $V_0 = 4.4898$, $V_{0\mathrm{Kap}} = 1.8449$  \end{tabular}      &    716.885          \\
NGC 2403 & NFW     &   $\rho_s  =  4.444 \cdot 10^7$, $r_s = 5.70$            &     615.187     \\
NGC 2403 & Einasto &  $\rho_{-2} = 9.122 \cdot 10^6 $, $r_{-2} = 6.27$, $\alpha = 0.217$                   &    587.506           \\
NGC 2403 & Burkert &  $\rho_0 = 2.943 \cdot 10^8 $, $r_0 = 2.21$         &  4918.37
\\ \hline
UGC 08490 & Our model    & \begin{tabular}[c]{@{}c@{}}$ \rho_{\text{core}} = 3.19368$, $r_c = 0.492455$, \\ $A = 1.3794$, $V_0 = 5.1724$, $V_{0\mathrm{Kap}} = 8.4615$, \end{tabular}                    &    26.6488         \\
UGC 08490 & NFW     &    $\rho_s  =  8.65038 \cdot 10^7$, $r_s = 2.51758$              &    3.54876          \\
UGC 08490 & Einasto &  $\rho_{-2} = 2.255 \cdot 10^7 $, $r_{-2} = 2.49$, $\alpha = 0.280$                &  2.79813        \\
UGC 08490 & Burkert &  $\rho_0 = 4.31961 \cdot 10^{8} $, $r_0 = 1.16206$          &  24.9981
\\ \hline
\end{tabular}
\label{table1}
\end{table}

\section{Discussions} \label{discussions}

In \figref{galaxies1} and \figref{galaxies2}, we compare the radial velocity profile of the Kapitza core-tail model developed in this work with observational data and with classical profiles widely used in the literature, simultaneously analyzing the rotation curves and density profiles.

In \figref{galaxies1}, we group the galaxies for which the proposed model provides the best performance in fitting the rotation curves, based on the $\chi^2$ values presented in Table \ref{table1}. The galaxies NGC 2903 and NGC 3198 are classified as barred spiral galaxies, while UGC 11914 and ESO 563-G021 are spiral galaxies. For this set of objects, the core-tail model outperforms the other density profiles considered.

In \figref{galaxies2}, we group the galaxies for which the fit obtained with the proposed model is not the best among all profiles, but still remains competitive. For the irregular galaxy NGC 2366, our model yields a fit comparable to the Einasto profile and superior to the NFW and Burkert profiles. In the case of NGC 2403, a spiral galaxy, the proposed model outperforms only the Burkert profile. Finally, for the spiral galaxy UGC 08490, the fit obtained with the proposed model is worse than those of the other profiles analyzed. Basic properties and morphological classifications of the galaxies were obtained from the NASA/IPAC Extragalactic Database (NED) \cite{NED}.

Furthermore, in the inner region, the resulting density profiles exhibit cored behavior, consistent with the soliton solution expected for dark matter condensates. This result agrees with models based on the Schrödinger-Poisson equations, frequently cited as potential alternatives to solve the cusp-core problem, since they provide halo centers with approximately constant density.

We also find that the numerical solutions exhibited oscillations in the tail region, whose origin is associated with the values assigned to the parameters $A$ and $V_0$. These unexpected oscillations are independent on the Kapitza potential and, to the best of our knowledge, their physical origin is not explicitly discussed in the existing literature. This fact suggests that the mathematical structure of the tail's system may induce intrinsic oscillatory regimes, whose physical interpretation still requires further investigation.

\begin{figure}[!htb]
\centering
\includegraphics[width=.48\textwidth]{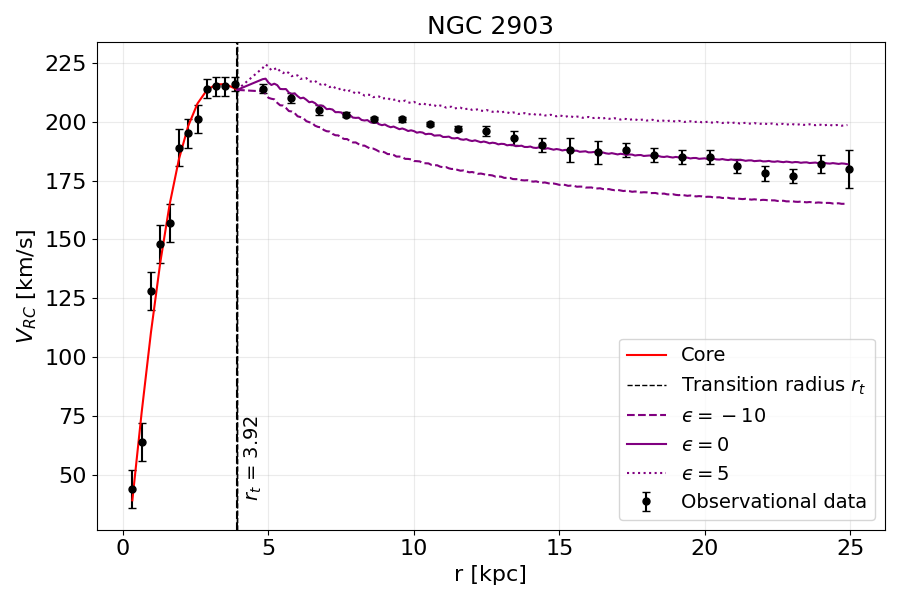}   % end subfigure
\quad 
\includegraphics[width=.48\textwidth]{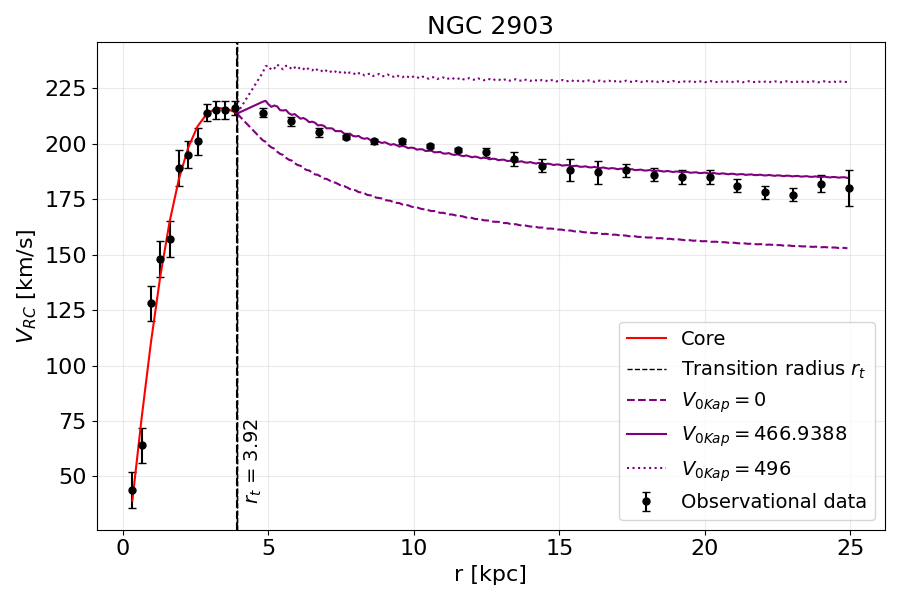}
\caption{Rotation curve of the galaxy NGC 2903 fitted by the core-tail model. The part $r < r_t$ corresponds to the solitonic core, while $r \geq r_t$ corresponds to the numerical tail solution, shown for different values of $\epsilon$ in the left panel and for different values of $V_{0\mathrm{Kap}}$ in the right panel. The dashed vertical line indicates the transition radius $r_t$. In both panels, the remaining reference values for the parameters are those presented in Table \ref{table1}. In each isolated variation, whether of $\epsilon$ or $V_{0\mathrm{Kap}}$, all other parameters remain fixed at the values in the table.}
\label{2903}
\end{figure}

In \figref{2903}, we show the rotation curves of the galaxy NGC 2903, highlighting the contribution of the solitonic core and the behavior of the tail region for different values of the interaction parameter $\epsilon$ in the left panel and $V_{0\mathrm{Kap}}$ in the right panel. It clearly shows how the tail solution of the core-tail model responds to variations in the BEC parameter $\epsilon$ (left panel) and in the Kapitza potential magnitude parameter $V_{0\mathrm{Kap}}$ (right panel). In the left panel, different values of the self-interaction parameter $\epsilon$ directly alter the effective lift of the halo in the outer region. We saw that positive values increase the orbital velocity after the transition radius $r_t$, producing a more sustained curve, while negative values reduce this lift and lead to a steeper decline. In the right panel, a qualitatively similar response is observed when varying the amplitude of the Kapitza potential $V_{0\mathrm{Kap}}$.  Increasing $V_{0\mathrm{Kap}}$ results in a more sustained tail with higher rotational velocities, while smaller values lead to a decrease. This confirms that the Kapitza potential provides an additional mechanism for regulating the halo dynamics in the tail region.

Therefore, from the results presented in Table \ref{table1}, we found that the model shows competitive performance in relation to the NFW, Einasto, and Burkert profiles. In several galaxies, the fit was elaborate or even superior to these empirical models, establishing that the combination of a solitonic core and a dynamic tail constitutes a consistent description for galactic halos.

\section{Conclusion} \label{conclusion}
This paper was based on a theoretical study of dark matter halos modeled as Bose-Einstein condensates, which are subject to an extra Kapitza type potential as recently realized in laboratory. We use computational tools and the hydrodynamic formulation of the Gross-Pitaevskii equation, coupled with the Poisson equation, to investigate the effect of this potential on the core-tail structure and the behavior of the resulting rotation curves. Our analysis is, in principle, qualitative, but also quantitative for specific cases, focusing on the influence of model parameters on halo morphology and solutions to the associated dynamic system.

We show that the core-tail model satisfactorily reproduces the profile presented in several galaxies, achieving competitive performance in relation to classical empirical profiles, such as NFW, Einasto, and Burkert, always adopting the minimum disk approach. The Kapitza potential plays a relevant role by introducing an additional mechanism capable of sensitively modulating the shape of the tail, influencing the radial decay of density and the scale of orbital velocities.

The model still involves questions that require further investigation. Among them, we highlight the need to understand the origin and physical interpretation of the Kapitza potential in astrophysical contexts, as well as to analyze the stability of the oscillations induced in the tail density. Future studies that systematically explore the parameter space and perform direct comparisons with fully numerical simulations may clarify these limitations and better delineate the regime of applicability of the model. Within these limitations, however, the results obtained here indicate that the introduction of the Kapitza potential significantly expands the set of possible solutions in Bose-Einstein condensate models for dark matter, providing a promising alternative for describing galactic halos with a core-tail structure.

\section*{Acknowledgements}
This work was supported by the Research Support Foundation of the State of Minas Gerais (FAPEMIG), the Research Support Foundation of the State of Espírito Santo (FAPES), and the Brazilian federal agencies CNPq and CAPES. The authors thank Iván Álvarez for helpful comments and clarifications.

%\bibliography{biblio}

\end{document}